\documentclass[twocolumn,showpacs,preprintnumbers,amsmath,amssymb,prl,superscriptaddress,letterpaper]{revtex4}

\usepackage{graphicx}% Include figure files
\usepackage{dcolumn}% Align table columns on decimal point
\usepackage{bm}% bold math
\usepackage{txfonts}

\begin{document}

%\preprint{APS/123-QED}

\title{Spatially resolved observation of dipole-dipole interaction between Rydberg atoms}

\author{C.~S.~E. van Ditzhuijzen}
 \affiliation{Van der Waals-Zeeman Institute, University of
Amsterdam, Valckenierstraat 65, 1018 XE Amsterdam, NL}
\author{A.~F. Koenderink}
\affiliation{FOM-Institute for Atomic and Molecular Physics,
Kruislaan 407, 1098 SJ Amsterdam, The Netherlands}
\author{J.~V. Hern\'andez}
\affiliation{Department of Physics, Auburn University, Alabama
36849-5311, USA}
\author{F. Robicheaux}
\affiliation{Department of Physics, Auburn University, Alabama
36849-5311, USA}
\author{L.~D. Noordam}
\affiliation{Van der Waals-Zeeman Institute, University of
Amsterdam, Valckenierstraat 65, 1018 XE Amsterdam, NL}
\author{H.~B. van Linden van den Heuvell}
 \email{heuvell@science.uva.nl}
\affiliation{Van der Waals-Zeeman Institute, University of
Amsterdam, Valckenierstraat 65, 1018 XE Amsterdam, NL}

\date{\today}

\begin{abstract}
We have observed resonant energy transfer between cold Rydberg
atoms in spatially separated cylinders. Resonant dipole-dipole
coupling excites the 49$\mathrm{s}$ atoms in one cylinder to the
49$\mathrm{p}$ state while the 41$\mathrm{d}$ atoms in the second
cylinder are transferred down to the 42$\mathrm{p}$ state. We have
measured the production of the 49$\mathrm{p}$ state as a function
of separation of the cylinders (0~-~80~$\mu$m) and the interaction
time (0~-~25~$\mu$s). In addition we measured the width of the
electric field resonances. A full many-body quantum calculation
reproduces the main features of the experiments.
\end{abstract}


\maketitle

Transport of excitations by resonant interaction between dipoles
is an ubiquitous phenomenon that is fundamental to a broad range
of disciplines, ranging from life sciences to quantum computing.
In biological systems resonant dipole-dipole interactions mediate
the ultrafast energy flow in light harvesting complexes
responsible for photosynthesis
\cite{Science.285.400,Nature.417.533}. Dipole-dipole interactions
between fluorophores, as first described by F\"orster
\cite{AnnPhys.2.55}, are now a workhorse tool in biological
imaging to measure nanoscale distances \cite{NatBioTech.21.1387}.
Furthermore, manipulating the coupling between dipoles is
essential to a diverse range of emerging fields. In nanophotonics
coherent dipole-dipole coupling between carefully placed
polarizable plasmonic nanoparticles is pursued as a tool to create
ultrasmall optical circuits
\cite{OptLett.23.1331,PRB.62.16356,JPCB.109.15808,PRB.74.033402}.
For realizing quantum computing proposals, understanding of, and
full control over, both the dynamics and the spatial properties of
coherent excitation transfer between quantum systems is a crucial
step
\cite{PRL.85.2208,PRL.87.037901,PRA.70.042703,PRA.72.042302,PRB.74.041307}.

The range of length and time scales over which dipole-dipole
coupling occurs is ultimately set by the magnitude of the dipole
moments of the involved atoms, molecules or quantum dots.
Unfortunately the resulting time scales for dipole coupling are
ultrafast, while the length scales are very small. Resolving and
manipulating the interaction both in time and space simultaneously
is therefore extremely challenging. For instance, temporal quantum
control over dipole-dipole coupling has been demonstrated on a
femtosecond timescale, yet without spatial information on
excitation transport \cite{Nature.417.533}. Conversely, Hettich et
al.\ have resolved coherent coupling over nanometer length scales
using spectral properties of two resonant molecules. Inherently
this spectroscopic experiment provided no insight in or control
over the temporal dynamics \cite{Science.298.385}. Already for
some time, it has been realized that the large dipole moments of
Rydberg atoms promise to remove this limitation. Interactions
between Rydberg atoms occur over length and time scales that are
easily addressable in an experiment \cite{PRA.70.042703}.
Observations of dipole-dipole coupling between Rydberg atoms so
far have only probed spectroscopic and dynamical properties
without direct control over the interatomic distances
\cite{PRL.80.249,PRL.93.233001,PRL.93.153001,PRA.73.032725,PRL.94.173001,PRL.97.083003,EPJD.40.37}.

\begin{figure}[hb]
    \includegraphics[width=.49\textwidth]{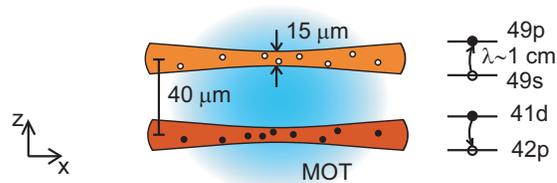}
    \caption{(color online) Two pulsed laser beams excite cold atoms
    to Rydberg states. The 41$\mathrm{d}$ atoms in one volume interact with
    the 49$\mathrm{s}$ atoms in the other volume. The resonant dipole-dipole
    interaction giving rise to a change of states $41\mathrm{d} + 49\mathrm{s} \rightarrow 42\mathrm{p} + 49\mathrm{p}$
    is in the near field limit
    (R$\sim$40~$\mu$m~$\ll$~$\lambda$=0.9~cm).
    We resolve the interaction in time and space by varying the beam separation.
    }\label{fig:cartoon}
\end{figure}

In the experiment presented in this Letter, we control the spatial
separation between Rydberg atoms in two dimensions and study the
dynamics of resonant excitation transfer by dipole-dipole
interactions. Distinct Rydberg states are created from a
magneto-optical trap (MOT) of Rb atoms by pulsed excitation using
two independent focused laser beams (Fig.\ \ref{fig:cartoon}).
Dipole-dipole coupling causes a transition of a Rydberg atom
(49$\mathrm{s}$) in one volume to a higher state (49$\mathrm{p}$)
and a simultaneous transition of an atom (41$\mathrm{d}$) in the
second volume to a lower state (42$\mathrm{p}$). This transition
occurs over separations up to 50~$\mu$m, over 10~$\mu$s time
scales.

Dipole-dipole interactions in Rydberg atoms can be induced by
tuning two transitions into resonance using a static electric
field. A transition in rubidium which has particularly large
transition dipole moments and a resonance at a low electric field
($\sim$0.4~V/cm) is
\begin{equation}
41\mathrm{d}_{3/2} + 49\mathrm{s}_{1/2} \leftrightarrow
42\mathrm{p}_{1/2} + 49\mathrm{p}_{3/2}\label{eq:reaction}
\end{equation}
The resonant transition frequency is 33~GHz. The strength of the
dipole-dipole interaction is given by
\begin{equation}
    V = \frac{\mathbf{\mu_1 \cdot \mu_2} - 3 (\mathbf{\mu_1 \cdot \hat R})
    (\mathbf{\mu_2 \cdot \hat R}) }{R^3}\label{eq:pot}
\end{equation}
where $\mathbf{R}$ is the distance vector between the interacting
particles, $\mathbf{\mu_1}$($\mathbf{\mu_2}$) is the dipole moment
of the $41\mathrm{d} \rightarrow 42\mathrm{p}$ ($49\mathrm{s}
\rightarrow 49\mathrm{p}$) transition. Both dipole moments are of
the order of 1000$a_0e$. The transition probability
$P\propto\sin^2(V t)$ shows quantum beats with a period of
$\sim$14$\mu$s for $R$=40$\mu$m (if $\mathbf{\mu_1}$,
$\mathbf{\mu_2} \parallel \mathbf{R}$). For short times ($V t
\ll$1), we approximate $P \propto t^2/R^6$. For dipoles on
parallel lines at distance $d$, the short-time probability
averages to $P \propto t^2/d^5$.

The setup used is similar to \cite{EPJD.40.13} and consists of a
standard $^{85}$Rb MOT. The atoms hardly move on the timescale of
the experiment as the average speed is below 0.3~$\mu$m$/\mu$s
(T=300~$\mu$K or less). The cold ground-state atoms (5s) are
excited to a Rydberg state by an 8~ns laser pulse of 594 nm in a
two-photon process. Two separate lasers, one for each of the two
Rydberg states, are focused next to each other in the MOT cloud.
The one that excites to the 49$\mathrm{s}$ state has a linewidth
of 0.09(1)~cm$^{-1}$ and a pulse energy of 3.0(1)~$\mu$J. For the
41$\mathrm{d}$ laser we have 0.21(2)~cm$^{-1}$ and 5.0(1)~$\mu$J.
The foci are imaged on a CCD camera, with a pixel size of
5.6~$\mu$m. The 49$\mathrm{s}$ beam can be laterally moved by a
motorized lens.

An electric field surrounding the MOT cloud is created by applying
a differential voltage on two 5.5~cm diameter circular plates,
2.5~cm apart, perforated by a 14~mm hole to allow MOT beams and
ionization products to pass. This electric field, parallel to the
laser beam separation and polarization, brings the dipole-dipole
transfer [Eq.\ \ref{eq:reaction}] into resonance. After some
interaction time an electric field pulse (from 0 to 150~V/cm in
5~$\mu$s) ionizes the atoms and the released electrons are
detected by a microchannel plate (MCP). This gives different
electron arrival times for the 49$\mathrm{p}$ and 49$\mathrm{s}$
atoms; the 41$\mathrm{d}$ and the 42$\mathrm{p}$ states are not
distinguished. We use the 49$\mathrm{p}$ signal as a measure of
the dipole-dipole interaction. Every electron is timed separately
(the detection efficiency for MCP's is typically 40-70\%) and all
data are averaged over 200 laser shots.

The initial number of measured 49$\mathrm{s}_{1/2}$ atoms per shot
is 12.5(2.5) and the total number of atoms in the 41$\mathrm{d}$
volume is 70(10), not taking into account the MCP's finite
detection efficiency. Of these atoms 15(3) are in the
43$\mathrm{p}$ state and we presume the same number in the
42$\mathrm{p}$ state. We get 24(5) atoms in the
41$\mathrm{d}_{5/2}$ state and 16(3) atoms in the relevant
41$\mathrm{d}_{3/2}$ state, based on the ratio given by our laser
polarization. During the experiment, the number of Rydberg atoms
slightly decreases due to spontaneous decay (rates around 8~kHz).
Second, black body radiation induces transitions to nearby lying
Rydberg states with rates around 10~kHz \cite{Tombook}. Reference
measurements with an isolated 41$\mathrm{d}$ beam result in 0.5
49$\mathrm{s}$ atom and 0.5 49$\mathrm{p}$ atom and are subtracted
from the data.

\begin{figure}[hb]
    \includegraphics[width=.495\textwidth]{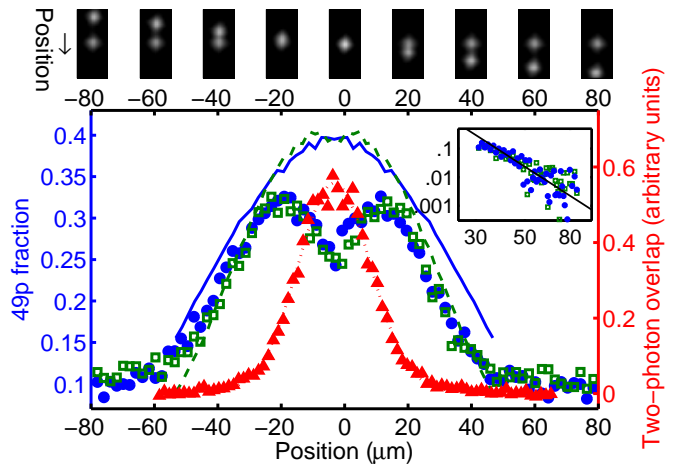}
    \caption{(color online) The 49$\mathrm{p}$ fraction after 10~$\mu$s
    as a function of the
position of the 49$\mathrm{s}$ beam at the resonance F$_1$ (blue
$\medbullet$) and F$_2$ (green $\square$). Blue solid (F$_1$) and
green dashed (F$_2$) lines are simulations. The measured beam
overlap (red $\blacktriangle$) is fitted to a Gaussian (red dotted
line). CCD pictures of the laser foci above the graph correspond
to the horizontal axis tick marks. The log-log plot inset depicts
the wings of the data together with a fit to
$d^{-5}$.}\label{fig:pos}
\end{figure}

The Rydberg volumes are cigarlike ellipsoids. Their length is
determined by the size of the MOT cloud ($\sim$0.5~mm). For
unsaturated excitation we expect that the diameter of the Rydberg
volume is $1/\sqrt{2}$ times the laser beam waist, because of the
two-photon process. However, laser fluence dependant measurements
show that both excitations are slightly saturated in our
experiment, resulting in slightly larger diameters. The laser
waists are determined by a two-photon overlap measurement. For
this we detune the 49$\mathrm{s}$ laser by 20~GHz to the blue and
we make sure that the laser pulses overlap in time, so that
absorbing one photon of each laser leads to excitation to the
44$\mathrm{d}$ state. By moving the 49$\mathrm{s}$ beam over the
41$\mathrm{d}$ beam one obtains the convolution of the two laser
beams in the 44$\mathrm{d}$ signal. A measurement is depicted in
Fig.\ \ref{fig:pos} and fitted to a Gaussian profile with a
1/$\sqrt{e}$ full width of 22.8(6)~$\mu$m. Measurements as a
function of saturation (not shown) yield a laser waist of
13.7(4)~$\mu$m for each laser, assuming identical beams. Taking
the measured saturation with increasing power into account, the
Rydberg volume diameters in our experiment are 11.6(4)~$\mu$m for
49$\mathrm{s}$ and 16.3(5)~$\mu$m for 41$\mathrm{d}$. Without
assuming identical beams, the convolution of the two saturated
Rydberg volumes (relevant for the experiment) has a full width of
less than 23(1)~$\mu$m. Due to the finite temperature, the
diameter of the cylinders increases at most 2~$\mu$m in 25~$\mu$s.

For the interpretation of the data, we simulated the populations
in each state by performing a fully quantum calculation with a
limited number of atoms fixed in space. The matrix elements were
computed using standard angular momentum algebra and numerical
integration for the radial matrix elements. We used the energy
levels of Ref.~\cite{PRA.67.052502} to determine the radial
functions needed for the matrix elements. No adjustments were made
to get better agreement with the measurements. We randomly placed
25 atoms in each cigar-shaped ellipsoid (Gaussian in 3D) and
performed a series of calculations starting with one randomly
picked 49$\mathrm{s}$ atom. We added neighboring atoms in the
calculation until the time dependent probability for the
$49\mathrm{s}\rightarrow49\mathrm{p}$ transition converged. We
obtained good results with two 49$\mathrm{s}$ atoms and the two
nearest 41$\mathrm{d}$ atoms, which are depicted in the figures;
full convergence was obtained with one additional 41$\mathrm{d}$
atom. This demonstrates that many-body effects are important.
Adding more electronic energy levels per atom or adding atoms
initially in the 42$\mathrm{p}$ or 43$\mathrm{p}$ state slightly
slowed down the transition times ($\sim$20\%).

\begin{figure}[hb]
    \includegraphics[width=.495\textwidth]{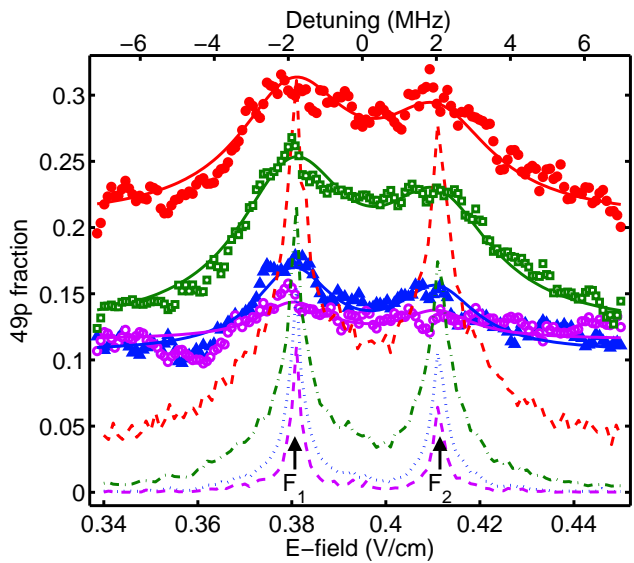}
    \caption{
(color online) The 49$\mathrm{p}$ fraction after 10~$\mu$s as a
function of electric field for different beam separations: red
$\medbullet$~21, green $\square$~31, blue $\blacktriangle$~41 and
purple $\medcirc$~51~$\mu$m. Solid lines are Lorentzian fits. Red
dashed (20~$\mu$m), green dash-dotted (30~$\mu$m), blue dotted
(40~$\mu$m) and purple dashed (50~$\mu$m) lines are simulations.
F$_1$ (0.38 V/cm) and F$_2$ (0.41 V/cm) are the resonances. The
top axis depicts a relative energy scale.}\label{fig:field}
\end{figure}

As a first experiment we monitored the 49$\mathrm{p}$ fraction
$N_{49\mathrm{p}}/(N_{49\mathrm{s}}+N_{49\mathrm{p}})$, resulting
from the resonant dipole-dipole interaction, after 10~$\mu$s as a
function of the applied static electric field. The measurements
performed with beam separations between 21 and 51~$\mu$m are
depicted in Fig.\ \ref{fig:field}. Two resonances can be seen, due
to a small difference in the Stark shift of the
49$\mathrm{p}_{3/2}$ $|m_j|$ states. At the field F$_1$(F$_2$),
$|m_j|=\frac{1}{2}$($\frac{3}{2}$) is resonant. The
41$\mathrm{d}_{3/2}$ splitting is not visible, because only
$|m_j|=\frac{1}{2}$ is excited by the laser. The field values have
a systematic error of 2\%, due to the uncertainty in the effective
plate distance. The calculated resonances fit within this error,
confirming that the creation of 49$\mathrm{p}$ in the
49$\mathrm{s}$ volume requires the tuning into resonance with the
transition in the adjacent 42$\mathrm{p}$ volume. For the top axis
a conversion of 127~MHz/(V/cm) is used, based on the calculated
difference of the total polarizabilities of the initial and final
states.

\begin{figure}[hb]
    \includegraphics[width=.495\textwidth]{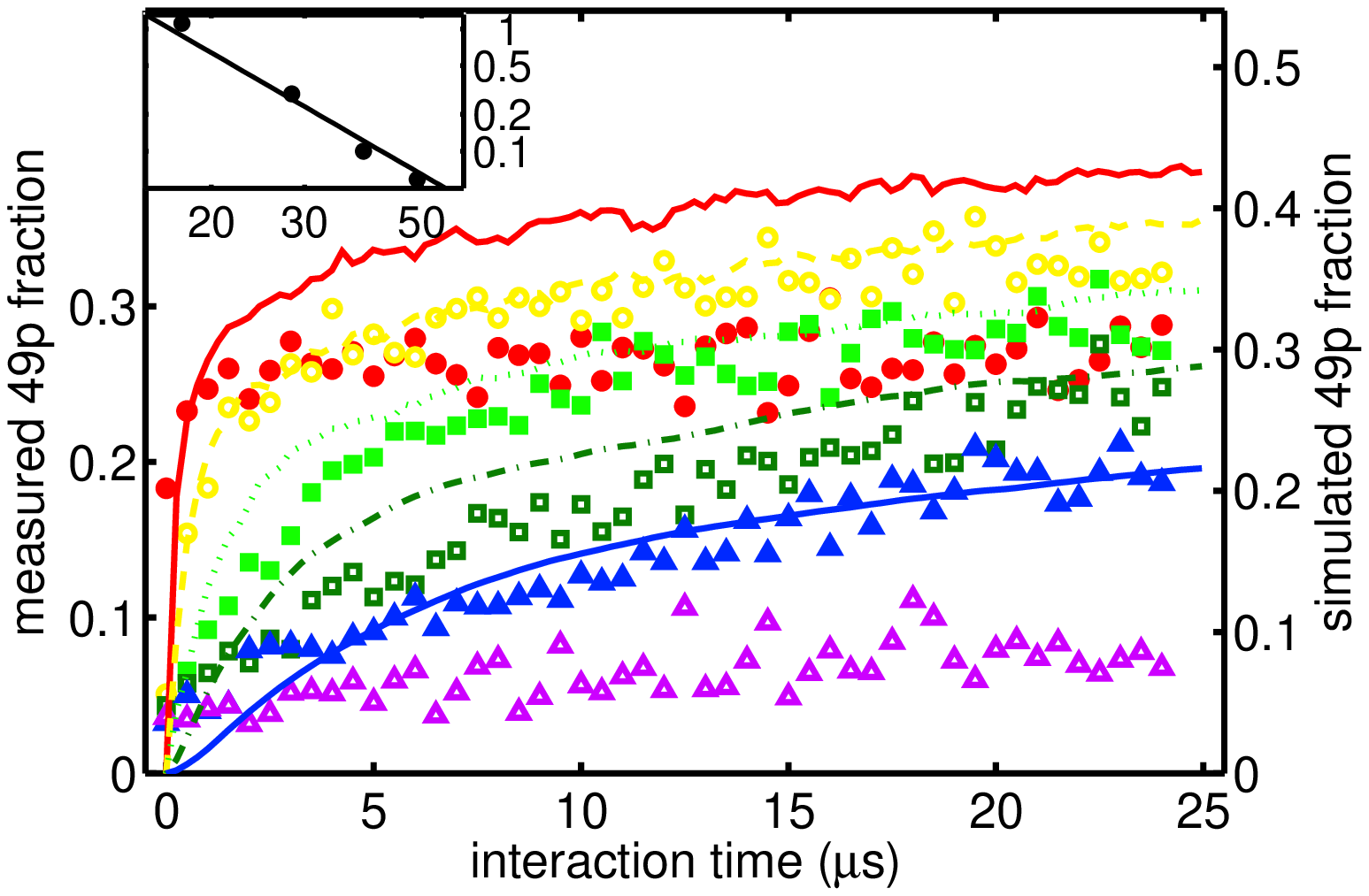}
    \caption{(color online) The growth of the 49$\mathrm{p}$ fraction
    for different beam
separations, measured data are red $\medbullet$ 0, yellow
$\medcirc$ 20, green $\blacksquare$ 30, dark green $\square$ 40,
blue $\blacktriangle$ 50~$\mu$m and purple $\triangle$ with only
49$\mathrm{s}$ atoms created (or $\infty$ separation). The
interaction time is controlled by varying the delay between the
laser excitation and the field ionization ramp. The simulations
are depicted as solid red (0~$\mu$m), dashed yellow (20~$\mu$m),
dotted green (30~$\mu$m), dash-dot dark green (40~$\mu$m) and
solid blue (50~$\mu$m) lines. The left axis refers to the measured
data, the right axis to the simulations. The inset show a log-log
plot of the transfer rate $1/\tau$ in MHz versus effective
distance together with a fit to $d^{-5/2}$.} \label{fig:time}
\end{figure}

It is clearly visible in Fig.\ \ref{fig:field} that the resonance
peaks become higher and broader as the distance between the foci
is reduced (FWHM are 18(2), 20(1), 29(1) and 29(1) mV/cm). This is
consistent with the notion that the dipole-dipole interaction
[Eq.\ (\ref{eq:pot})] gets stronger with shorter distance. These
features are also reproduced in the simulations. However, the
widths of the calculated resonance peaks are much narrower than
observed in the experiment. This is mainly due to the magnetic
field of the MOT ($\sim$1.4~MHz). It will further be discussed in
connection with Fig.\ \ref{fig:time}, as well as the origin of the
background of 0.1. The increasing background at shorter distance,
visible in simulation and measurement, is due to an increase of
overlap between the Rydberg volumes, as well as presumably
many-body effects.

To investigate the distance dependance of the interaction, we
tuned the field to each of the resonances and measured the
transferred fraction after 10~$\mu$s as a function of the
separation (Fig.\ \ref{fig:pos}). The most important result in
this figure is that a range of distances exists, where the overlap
of the lasers vanishes, while the interaction is still clearly
present. This demonstrates dipole-dipole energy transfer between
Rydberg atoms in separate volumes.

Figure \ref{fig:pos} also shows simulations of the experiment. The
effective range of interaction is well reconstructed, but slightly
overestimated in the calculation. The inset shows that the wings
of the data follow a $d^{-5}$ behavior, as expected. A striking
discrepancy with the simulation is that the experimental signal
decreases where the beams overlap. This is because nonresonant
processes take place in the high density 41$\mathrm{d}$ beam,
reducing the number of 49$\mathrm{s}$ atoms from 12.5(2.5) to
10(2); we particularly observe 4(1) atoms in the 47$\mathrm{p}$
state, but possibly Penning ionization occurs as well
\cite{PRA.69.063405}. Besides, there are 5(1) 44$\mathrm{d}$ atoms
due to the two-laser excitation. Within the accuracy of the
experiment there is no significant difference between the
resonance F$_1$ (blue $\medbullet$) and the resonance F$_2$ (green
$\square$). The asymmetry of the measured curve might be due to a
deviation from a Gaussian of the laser beam profiles or a decrease
in the density for positive positions ($\sim$10\%).

Next we will discuss the main result in this Letter: the temporal
evolution of the interaction. In Fig.\ \ref{fig:time} the
evolution of the 49$\mathrm{p}$ fraction is measured for various
beam separations with the field tuned to the resonance F$_1$. The
figure clearly shows that the rate of transfer slows down for
larger distance, but even at a 50~$\mu$m separation an appreciable
transfer is observed. The rates slow down because the interaction
strength decreases strongly with distance [Eq.\ (\ref{eq:pot})].
Quantum beat oscillations, which are expected on the basis of the
coherent coupling between atoms, are not observed. A spread in
distances of the interacting atoms, and hence a spread in
interaction strength, causes dephasing of these oscillations. The
transition rate $1/\tau$ in the inset is based on the time $\tau$
where the measured 49$\mathrm{p}$ fraction has reached 0.17. It is
plotted versus distance, slightly corrected for the finite
cylinder width. The data points nicely fit the scaling law
$d^{-5/2}$. The minor deviation might be related to the different
asymptotes of the curves.

Figure \ref{fig:time} also shows results of the simulations. The
curves reproduce the experimental data in a qualitative way, but
the calculated growth rate is slightly higher and the final
production is somewhat different. An important factor to note is
that temporal fluctuations in the electric field, that are fast
with respect to the dipole-dipole interaction rate, will slow down
the transfer rate. These fluctuations diabatically detune an
interacting atom pair in and out of resonance, thereby reducing
the fraction of time during which the interaction actually takes
place. Indeed, in early runs of the experiment, where the
high-frequency noise level of the field generated by the field
plates was 10 rather than 2~mV/cm, the observed transfer times
were a factor 2.8(5) longer. Moving ions in the MOT cloud might
also contribute to electrical field noise. Furthermore, the
initially present 42$\mathrm{p}$ and 43$\mathrm{p}$ atoms also
slow down the transfer rate, as indicated by the simulations.

The final 49$\mathrm{p}$ fraction is not 50\%, which one would
expect on the basis of binary interaction. This reduced transfer
is due to time-independent broadening effects, mainly the magnetic
field (see also Fig.\ \ref{fig:field}). However, since the
transferred fraction is also below 50\% in the simulations, it is
probable that also many-body effects play a role. The main process
can be accompanied by interactions between the initial
49$\mathrm{s}$ and the reaction product 49$\mathrm{p}$, as well as
between 41$\mathrm{d}$ and 42$\mathrm{p}$. These processes are
always resonant and occur within a single cylinder. Spurious
processes within the 49$\mathrm{s}$ beam can be observed when the
41$\mathrm{d}$ laser is turned off, depicted as purple $\triangle$
in Fig.\ \ref{fig:time}. This is mainly due to black body
radiation; the transition from 49$\mathrm{s}$ to 49$\mathrm{p}$
has a rate of 6.7~kHz \cite{Tombook}, but possibly also
nonresonant two-body processes play a role.

Up till now distance dependence of dipole-dipole interactions in
Rydberg atoms has only been measured indirectly by varying the
density. Here, we have taken the next step by having interaction
between atoms in different volumes at a well defined separation.
This is remarkable since an interaction is measured between two
mesoscopic gaseous systems at a macroscopic distance. We map the
interaction by varying the electric field, mutual separation and
interaction time and our data fit the here presented simulations
as well as a straightforward scaling law. The distance restriction
is not strong enough to show the coherence of the underlying
processes. Furthermore, due to the fact that only interactions at
large distances play a role, this experiment is more sensitive to
many-body interactions and external fields. When this approach is
combined with position-sensitive measurements \cite{EPJD.40.13} it
allows for many variations with more complexity and/or reduced
dimensionality and hence coherent evolution.

We thank T.\ F.\ Gallagher for fruitful discussions and A.\ F.\
Tauschinsky for experimental contributions. This work was funded
by FOM,  NWO and NSF (Grant No.\ 0355039).

\end{document}